\documentclass[a4paper,12pt]{article}
\usepackage[pctex32]{graphics}

\begin{document}
\topmargin 0pt
\oddsidemargin 0mm

\newcommand{\mn}{\mu\nu}
\newcommand{\bta}{\beta}
\newcommand{\gmm}{\gamma}
\newcommand{\del}{\delta}
\newcommand{\omg}{\omega}
\newcommand{\sgm}{\sigma}
\newcommand{\lmd}{\lambda}
\newcommand{\tha}{\theta}
\newcommand{\vph}{\varphi}
\newcommand{\Alp}{\Alpha}
\newcommand{\Bta}{\Beta}
\newcommand{\Gmm}{\Gamma}
\newcommand{\Del}{\Delta}
\newcommand{\Omg}{\Omega}
\newcommand{\Sgm}{\Sigma}
\newcommand{\Lmd}{\Lambda}
\newcommand{\Tha}{\Theta}
\newcommand{\half}{\frac{1}{2}}
\newcommand{\rnd}{\partial}
\newcommand{\nab}{\nabla}
\newcommand{\fr}{\frac}

\newcommand{\beq}{\begin{equation}}
\newcommand{\eeq}{\end{equation}}
\newcommand{\beqa}{\begin{eqnarray}}
\newcommand{\eeqa}{\end{eqnarray}}

\begin{titlepage}

\vspace{5mm}
\begin{center}
{\Large \bf  Critical gravity as van Dam-Veltman-Zakharov
discontinuity in anti de Sitter space } \vspace{12mm}

{\large   Yun Soo Myung \footnote{e-mail
 address: ysmyung@inje.ac.kr}}
 \\
\vspace{10mm} {\em  Institute of Basic Science and School of
Computer Aided Science, Inje University, Gimhae 621-749, Korea }

\end{center}

\vspace{5mm} \centerline{{\bf{Abstract}}}
 \vspace{5mm}
 We consider  critical gravity as van Dam-Vletman-Zakharov (vDVZ)
discontinuity in anti de Sitter space. For this purpose, we
introduce the higher curvature gravity.   This discontinuity can be
confirmed by calculating the residues of relevant  poles explicitly.
For the non-critical gravity of $0<m_2^2<-2\Lambda/3$, the scalar
residue of a massive pole is given by $2/3$ when taking the $\Lambda
\to 0$ limit first and then the $m^2_2 \to 0$ limit. This indicates
that the vDVZ discontinuity occurs in the higher curvature theory,
showing that propagating degrees of freedom is decreased from 5 to
3.  However, at the critical point of $m^2_2=-2\Lambda/3$, the
tensor residue of a massive pole blows up and scalar residue is
$-5/36$, showing the unpromising feature of the critical gravity.
\end{titlepage}
\newpage
\renewcommand{\thefootnote}{\arabic{footnote}}
\setcounter{footnote}{0} \setcounter{page}{2}

\section{Introduction}
There has been much interest in the massless limit of the massive
graviton propagator~\cite{Hig,KMP,POR,GNie,DW}. A key issue of this
approach is that van  Dam-Veltman-Zakharov (vDVZ)
discontinuity~\cite{DVZ} is peculiar to Minkowski space, but it
seems unlikely to arise in (anti) de Sitter space. The vDVZ
discontinuity implies that the limit of $M^2_{\rm MP} \to 0$ does
not yield a massless graviton at the tree level such that the
Einstein gravity (general relativity: GR) is isolated from the
massive gravity.   One has usually introduced the  Fierz-Pauli mass
term with mass squared $M^2_{\rm FP}$~\cite{FP} for this purpose.

If the cosmological constant (CC, $\Lambda$) was introduced and a
smooth $M^2_{\rm FP}/\Lambda \to0$  limit exists, first taking the
$M^2_{\rm FP} \to 0$ limit (and then, the $\Lambda \to 0$ limit)
recovers a massless graviton, leading to no vDVZ discontinuity in
the Einstein gravity. It is worth noting that $M^2_{\rm FP} \to0$
and $\Lambda \to 0$ limits do not commute. First taking the $\Lambda
\to 0$  limit, one encounters the vDVZ discontinuity~\cite{POR}.
Another resolution to the discontinuity is possible to occur even in
Minkowski space, if the Schwarzschild radius of the scattering
objects is taken  to be the second mass scale~\cite{Vain}. However,
these all belong to the linearized (tree) level calculations. If
one-loop graviton vacuum amplitude is computed  for a massive
graviton~\cite{DDLS}, the discontinuity appears again. This means
that the apparent absence of the vDVZ discontinuity may be
considered as  an artifact of the tree level approximation. Also,
there was  the Boulware-Deser instability which states that at the
non-linearized level, a ghost appears in the massive gravity
theory~\cite{BD}.

On the other hand,   critical gravities were recently investigated
in the AdS space as candidates of quantum
gravity~\cite{LP,DLLPST,AF,BHRT}. In the framework of  GR with
higher curvature terms (namely, higher curvature gravity), the
critical gravity is determined by two conditions of \beq \label{tcc}
 \alpha=-3\beta,~~\beta=-1/2\Lambda.
 \eeq
 At the critical point, the graviton tensor $h^{\rm cr}_{\mn}$ satisfies the fourth-order equation
 under the transverse and traceless gauge and its solution takes the log-form.
 The non-unitarity issue of log-gravity is not still resolved, indicating that the log-gravity
suffers from the ghost problem~~\cite{PR,LLL}. To resolve this ghost
problem, it was suggested that imposing unitarity may require
suppressing the log-modes by choosing an appropriate boundary
condition, leaving the Einstein gravity in the IR
region~\cite{LP,LPP}.

In this work, we wish to regard the appearance of  critical gravity
as the  vDVZ discontinuity of the  higher curvature gravity in the
AdS$_4$ space.  This picture may be clear when reminding  that the
critical gravity was originated from the chiral point in
cosmological topologically massive gravity in three dimensions: a
negative-energy massive graviton disappears at the chiral point of
$\mu=1/\ell$ and thus, it becomes a massless left-handed mode of the
3D Einstein gravity~\cite{LSS}. It is well known that the 3D
Einstein gravity is a gauge theory, which means that there is no
propagating degrees of freedom (DOF).
 This
cosmological topological massive gravity at the critical point
(CCTMG) may be described by using the logarithmic conformal field
theory (LCFT)~\cite{GJ,Myung} even for the zero central charge
$c_L=0$.   Another massive generalization of the 3D Einstein gravity
was proposed  by adding a specific quadratic curvature term to the
Einstein-Hilbert action~\cite{bht,bht2}. This gravity theory became
known as new massive gravity (NMG).  This theory was designed for
reproducing the ghost-free Fierz-Pauli action for a massive
propagating graviton in the linearized level.  In this sense, the
NMG is the non-linear realization of  a massive graviton. Unlike the
TMG, the NMG preserves parity.  As a result, the gravitons acquire
the same mass for both helicity states, indicating 2 DOF.  At the
critical point of $m^2=1/2\ell^2$, two massive modes turned out to
be massless left/right-handed modes  of the 3D Einstein gravity in
the AdS$_3$ space~\cite{LS}. Also, this corresponds to   zero
central charges $c_{L/R}=0$ on the boundary CFT.   We note that {\it
the critical gravity is just the higher dimensional extension of the
NMG at the critical point}.

It was argued that there is no vDVZ discontinuity in  GR with higher
curvature term (for example, $R-2\Lambda +\beta R^2$) in the AdS$_4$
space~\cite{Neu}.  Recently, a similar analysis including higher
curvature terms was performed in $D$-dimensional anti-de Sitter
 space, including the NMG~\cite{GT}. We comment that
the vDVZ discontinuity appeared in $f(R)$ gravity  because $f(R)$
gravity means GR with an additional scalar, implying 3 DOF
initially~\cite{Myungf}. In general, the vDVZ discontinuity was
closely related to the decreasing DOF of $5\to 3$. Hence, the
argument in Ref.\cite{Myungf} may be not justified because  unless
one includes $\alpha R_{\mu}R^{\mn}$, one might not see the vDVZ
discontinuity in the massless limit.

Reminding that the vDVZ discontinuity is the massless limit of the
massive gravity in the Minkowski space,   the critical gravity
appears as the massless limit of higher curvature gravity in the
AdS$_4$ space.  Therefore, we insist that the critical gravity may
be represented  as the vDVZ discontinuity of the  higher curvature
gravity in the AdS$_4$ space. In this case, we expect to have  a
decreasing DOF ($5 \to ?$)  at the critical point.

\section{Higher curvature gravity}

We start with  the  higher curvature  gravity  including a bare
cosmological constant $\Lambda_0$  \beq \label{act-0} I =
\frac{1}{2\kappa^2} \int d^4x \sqrt{- g} \Big \{ R -2\Lambda_0
+\beta R^2+\alpha R_{\mu\nu}R^{\mu\nu}\Big \} \label{Action} \eeq
with the gravitational constant $\kappa^2=8 \pi G$. Here we choose
$\beta$ and $\alpha$ to match with the convention in Ref.~\cite{LP}
and their mass dimensions are $[\alpha]=[\beta]=-2$ with $\alpha<0$
and $\beta>0$. Also, we follow the signature of $(-+++)$. In the
case of $\Lambda_0=0$, the theory is renormalizable  and it
describes  8 DOF (a massless spin-2 graviton with 2 DOF, a massive
spin-2 graviton with 5 DOF, and a massive scalar with 1
DOF)~\cite{stelle}. However, the massive graviton suffers from
having ghosts. We note again that $f(R)$ gravity with $\alpha=0$ has
3 DOF (a massless spin-2 graviton  and a massive scalar) without
ghost~\cite{Myungf}.

 The equation of motion is given by \beq \mbox{}
G_{\mn}+E_{\mn}=0, \label{FEQ} \eeq where the Einstein tensor
$G_{\mn}$ and $E_{\mn}$ are given by
\begin{eqnarray}
G_{\mn}&=&R_{\mn}-\frac{1}{2}Rg_{\mn}+\Lambda g_{\mn},\\
E_{\mn}&=&2\alpha\Big(R_{\mu\rho} R_\nu~^\rho-\frac{1}{4}
R_{\rho\sigma}R^{\rho\sigma}g_{\mn}\Big)+2\beta
R\Big(R_{\mn}-\frac{1}{4} Rg_{\mn}\Big) \nonumber \\
&+& \alpha\Big(\nabla^2R_{\mn}+\nabla_\rho \nabla_\sigma
R^{\rho\sigma}g_{\mn}-2\nabla_\rho
\nabla_{(\mu}R_{\nu)}~^\rho\Big)+2\beta
\Big(g_{\mn}\nabla^2-\nabla_\mu\nabla_\nu\Big)R.
\end{eqnarray}
 In
this case, the vacuum solution to (\ref{FEQ}) is
 the AdS$_4$  space whose geometry is expressed in
terms of the metric ($\bar g_{\mn})$  as \beq \bar
R_{\mu\nu\rho\sigma}=\frac{\Lambda}{3}(\bar g_{\mu\rho}\bar
g_{\nu\sigma}- \bar g_{\mu\sigma}\bar g_{\nu\rho}),~~~ \bar
R_{\mn}=\Lambda \bar g_{\mn},~~~ \bar R=4 \Lambda=-\fr{12}{\ell^2}.
\label{AdS} \eeq Its line element takes the form \beq ds^2_{\rm
AdS}=\bar{g}_{\mn} dx^\mu dx^\nu= -
\Big(1+\fr{r^2}{\ell^2}\Big)dt^2+\frac{dr^2}{\Big(1+\fr{r^2}{\ell^2}\Big)}+r^2d\Omega^2_2.
\eeq We note that the effective CC $\Lambda$ is equal to the bare CC
$\Lambda_0$ only in four dimensions.

 In order to study the propagation of the metric,
we introduce the perturbation around the AdS$_4$ space \beq
g_{\mu\nu} = \bar g_{\mu\nu} +h_{\mu\nu}. \label{Per} \eeq Hereafter
we denote the background value  with overbar ( $\bar{}$ ). We expect
that the theory describes  8 DOF in the  AdS$_4$ space, too. The
linearized equation to Eq.(\ref{FEQ}) with the external source
$T_{\mn}$ takes the form
\begin{eqnarray} \label{lineq}
&&\Bigg[1+\frac{4\alpha\Lambda}{3}+8\beta\Lambda \Bigg] \delta
G_{\mn}(h) +\alpha\Bigg[\bar{\nabla}^2 \delta
G_{\mn}(h)-\frac{2\Lambda}{3} \delta R(h)\bar{g}_{\mn}\Bigg]
\nonumber \\
&&+(\alpha+2\beta)\Big[\bar{g}_{\mn}\bar{\nabla}^2-\bar{\nabla}_\mu
\bar{\nabla}_\nu+\Lambda\bar{g}_{\mn}\Big]\delta R(h) = T_{\mn},
\label{PEQ} \end{eqnarray}  where the linearized Einstein tensor  is
given by~\cite{GT}   \beq \label{linEeq}\delta G_{\mn}(h)=\delta
R_{\mn}(h)-\frac{\bar{g}_{\mn}}{2}\delta R(h)-\Lambda h_{\mn}. \eeq
The linearized Ricci tensor and the linearized scalar curvature take
the forms, respectively,
\begin{eqnarray}
 \delta R_{\mn}(h)&=&\frac{1}{2}\Big[-\bar{\nabla}^2h_{\mn} -\bar{\nabla}_\mu\bar{\nabla}_\nu
h+2\bar{\nabla}^\rho \bar{\nabla}_{(\mu} h_{\nu)\rho}\Big], \nonumber \\
&=&\frac{1}{2}\Big[\Delta^{(2)}_L h_{\mn}-\bar{\nabla}_\mu
\nabla_\nu h+2\bar{\nabla}_{(\mu}\bar{\nabla}^\rho
h_{\nu)\rho}\Big],
\label{eqr} \\
 \delta R(h)&=& \bar{g}^{\mn}\delta R_{\mn}(h)- h^{\mn}\bar{R}_{\mn}=
\bar{\nabla}^\rho \bar{\nabla}^{\sigma}
h_{\rho\sigma}-\bar{\nabla}^2h-\bar{\Lambda}h, \label{eqs}
\end{eqnarray}
where the Lichnerowicz operator $\Delta^{(2)}_L h_{\mn}$ takes the
form \beq \label{Lich}\Delta^{(2)}_L h_{\mn}=-\bar{\nabla}^2h_{\mn}
+\frac{8\Lambda}{3}\Big(h_{\mn}-\frac{h}{4}\bar{g}_{\mn}\Big).\eeq
The trace of (\ref{lineq}) leads to \beq \label{traceq}
\Bigg[2(\alpha+3\beta)\bar{\nabla}^2- 1\Bigg]\delta R(h)=T.\eeq
Acting  $\nabla^\mu$ to (\ref{lineq}) leads to zero, indicating that
the Bianchi identity is satisfied  in the linearized level. As a
result, one finds the source-conservation law \beq \label{csc}
\bar{\nabla}^\mu T_{\mn}=0 \eeq which states that $T_{\mn}$ is
covariantly conserved with respect to the background metric of
AdS$_4$ space.

 At
this stage, we note that the linearized equation (\ref{lineq})
without external sources $T_{\mn}$  is invariant under linearized
diffeomorphisms as \beq \delta_\xi h_{\mn}=\bar{\nabla}_{\mu}
\xi_\nu+ \bar{\nabla}_{\nu} \xi_\mu, \eeq because of \beq
\delta_{\xi} \delta G_{\mn}(h)=0,~~~\delta_\xi \delta R(h)=0. \eeq
This implies that divergence and double divergence do not provide
any constraint on $h_{\mn}$. Hence, the gauge invariant (physical)
quantity is  left undetermined by the linearized equation
(\ref{lineq}). We use the diffeomorphisms to impose the transverse
gauge condition  \begin{equation} \label{gfix} \bar{\nabla}^\mu
h_{\mn}=\bar{\nabla}_\nu h.
 \end{equation}
Additionally,  its divergence is given by
 \beq \label{dgfix}
 \bar{\nabla}^\mu \bar{\nabla}^\nu
h_{\mn}=\bar{\nabla}^2 h. \eeq
 Using these conditions,
we rewrite the linearized Ricci tensor, scalar curvature,  and
Einstein tensors, respectively,  as
\begin{eqnarray}
&& \delta R_{\mn}(h)= \frac{1}{2}\Big[\Delta^{(2)}_L h_{\mn}
+\bar{\nabla}_\mu\bar{\nabla}_\nu h\Big],
\label{eqr1} \\
&& \delta R(h)=-\Lambda h, \label{eqs2} \\
&& \delta
 \label{eqs3}G_{\mn}(h)=\frac{1}{2}\Bigg[-\bar{\nabla}^2h_{\mn}+\bar{\nabla}_\mu
\bar{\nabla}_\nu h
+\frac{2\Lambda}{3}\Big(h_{\mn}+\frac{h}{2}\bar{g}_{\mn}\Big)\Bigg].
\end{eqnarray}

In order to find physically propagating modes, we decompose
 the metric perturbation $h_{\mn}$ with 10 DOF covariantly   into
\beq h_{\mn}=h^{TT}_{\mn}+\bar{\nabla}_{(\mu}
V_{\nu)}+\bar{\nabla}_{\mu} \bar{\nabla}_{\nu} \phi +\psi
\bar{g}_{\mn}, \eeq where $h^{TT}_{\mn}$ is the transverse traceless
(TT) tensor with 5 DOF ($\bar{\nabla}^\mu h^{TT}_{\mn}=0,h^{TT}=0$),
$V_\nu$ is a divergence free vector with 3 DOF ($\bar{\nabla}^\mu
V_\mu=0$), and $\phi$ and $\psi$ are scalar fields with 2 DOF. These
imply two relations \beq \label{relh}
\bar{\nabla}^2h=\bar{\nabla}^4\phi+4 \bar{\nabla}^2\psi,~~
\bar{\nabla}^\mu\bar{\nabla}^\nu h_{\mn}=\bar{\nabla}^4\phi+\Lambda
\bar{\nabla}^2\phi+ \bar{\nabla}^2\psi. \eeq  One-particle
scattering amplitude is mostly computed by choosing a condition of
(\ref{dgfix}) even if one does not impose the gauge condition
(\ref{gfix})~\cite{GT}. Then, considering (\ref{relh}) together with
this condition leads to \beq 3 \bar{\nabla}^2\psi=\Lambda
\bar{\nabla}^2\phi \eeq which implies that two scalars $\phi$ and
$\psi$ are not independent under the condition (\ref{dgfix}).
Plugging this into the first relation of (\ref{relh}), one finds a
relation between the trace $h$  of $h_{\mn}$ and scalar $\psi$ as
\beq \label{h-eq}
h=\frac{3}{\Lambda}\Big[\bar{\nabla}^2+\frac{4\Lambda}{3}\Big]\psi,
\eeq which implies that $\bar{\nabla}^2+\frac{4\Lambda}{3}$ is not a
physical pole and  thus, it belongs to a removable (unphysical)
pole~\cite{DDLS}.
 Using  (\ref{traceq}), (\ref{eqs2}),  and (\ref{h-eq}), we
express $\psi$ in terms of the trace $T$ of external sources
$T_{\mn}$ as \beq \psi=\frac{ 1}{3
\Big[-2(\alpha+3\beta)\bar{\nabla}^2+1
\Big]\Big(\bar{\nabla}^2+\frac{4}{3} \Lambda\Big)}T\eeq which means
that  $\psi$  becomes a  massive propagating scalar on AdS$_4$ space
unless $\alpha=-3\beta$. However, imposing this condition of
$\alpha=-3\beta$  to obtain critical gravity leads to that $\psi$
becomes trivial.

To find the transverse traceless part $h^{TT}_{\mn}$, we need to use
the
 Lichnerowicz operator $\Delta_L^{(2)}$ acting on spin-2 symmetric
 tensors  defined  in (\ref{Lich}).
Taking into account the TT condition, we rewrite the linearized
Einstein tensor as \beq \delta
G^{TT}_{\mn}(h)=\frac{1}{2}\Delta^{(2)}_Lh^{TT}_{\mn}-\Lambda
h^{TT}_{\mn}. \eeq  Then, we express $h_{\mn}^{TT}$ in terms of
external sources as \beq
h_{\mn}^{TT}=\frac{2}{\Big[1+(\frac{4\alpha}{3}+8\beta)\Lambda+\alpha
\bar{\nabla}^2\Big](\Delta_L^{(2)}-2\Lambda)}T^{TT}_{\mn}, \eeq
where the transverse traceless source ($ \bar{\nabla}^\mu
T_{\mn}^{TT}=0, T^{TT}=0$ with (\ref{csc})) is given by~\cite{POR}
\beq T_{\mn}^{TT}= T_{\mn} - \frac{1}{3} T \bar{g}_{\mn}
+\frac{1}{3}\Bigg[\frac{\bar{\nabla}_\mu \bar{\nabla}_\nu +
\frac{\Lambda}{3}\bar{g}_{\mn}}{\bar{\nabla}^2 + \frac{4
\Lambda}{3}}\Bigg]T. \eeq For the GR with CC, $h_{\mn}^{TT}$ takes a
simple form \beq
h_{\mn}^{TT}=\frac{2}{\Delta_L^{(2)}-2\Lambda}T^{TT}_{\mn}, \eeq
which for $\Lambda=0$ gives us the gravitational wave~\cite{FH} \beq
h_{\mn}^{TT}=-\frac{2}{\nabla^2}T^{TT}_{\mn} \eeq in the Minkowski
space.

 We are now in a position to define the tree-level
(one particle) exchange amplitude between two external sources
$\tilde{T}_{\mn}$ and $T_{\mn}$  as \beq A=\frac{1}{4} \int d^4x
\sqrt{-\bar{g}}\tilde{T}_{\mn}(x) h^{\mn}(x) \equiv
\frac{1}{4}\Big[\tilde{T}_{\mn} h^{TT \mn }+\tilde{T} \psi\Big],
\eeq where we suppress the integral to have a notational simplicity
in the last expression. Finally, the scattering amplitude takes the
form~\cite{GT} \beqa \label{opamp}4A
&=&2\tilde{T}_{\mn}\Big[\Big(1+(\frac{4\alpha}{3}+8\beta)\Lambda+\alpha
\bar{\nabla}^2\Big)(\Delta_L^{(2)}-2\Lambda)\Big]^{-1}T^{\mn}
\nonumber \\
 &+&\frac{2}{3}\tilde{T}
\Big[\Big(1+(\frac{4\alpha}{3}+8\beta)\Lambda+\alpha
\bar{\nabla}^2\Big)(\bar{\nabla}^2+2\Lambda)\Big]^{-1}T
\\ \nonumber
&-&\frac{2\Lambda}{9}\tilde{T}
\Big[\Big(1+(\frac{4\alpha}{3}+8\beta)\Lambda+\alpha
\bar{\nabla}^2\Big)(\bar{\nabla}^2+2\Lambda)\Big]^{-1}\Bigg[\bar{\nabla}^2+\frac{4\Lambda}{3}\Bigg]^{-1}T
\\ \nonumber
&+& \frac{1}{3}\tilde{T}\Bigg[-2(\alpha+3\beta)\bar{\nabla}^2+1
\Bigg]^{-1}\Bigg[\bar{\nabla}^2+\frac{4\Lambda}{3}\Bigg]^{-1}T.
\eeqa We note that there is no vector $V_\mu$ in the scattering
amplitude because it could be eliminated by choosing a further
gauge-fixing~\cite{DDLS}.

\section{vDVZ discontinuity in Minkowski space}

In the case of $\Lambda=0$, $m^2_0\not=0$, and $m^2_2\not=0$, the
one-particle scattering amplitude (\ref{opamp}) takes a simple form
in the momentum space ($\nabla^2\to -p^2$) as \beqa
\label{mamp}[4A(p)]_{\rm Min}&=&
\Bigg[\tilde{T}_{\mn}\frac{2}{p^2} T^{\mn}- \tilde{T}\frac{1}{p^2} T\Bigg] \nonumber \\
&-&\Bigg[\tilde{T}_{\mn}\frac{2}{p^2+m^2_2} T^{\mn}-\frac{2}{3}
\tilde{T}\frac{1}{p^2+m^2_2} T\Bigg] \nonumber
\\
&+& \frac{1}{3} \tilde{T} \frac{1}{p^2+m^2_0} T \eeqa where two
masses are given by \beq
m^2_0=\frac{1}{\alpha+3\beta},~~m^2_2=-\frac{1}{ \alpha}. \eeq In
the convention of  $\alpha<0$ and $\beta>0$, there is a massive
ghost with positive mass squared ($m^2_2>0$) because of negative
sign in the second line. Also, from the third line, one finds that a
massive scalar is free from  ghost and tachyon for $\alpha>-3\beta$.
The amplitude (\ref{mamp}) describes  8 DOF of 2 for massless
graviton, 5 for massive graviton (ghost), and 1 for massive
scalar~\cite{stelle}. In obtaining the second term $-1$ in the first
line, we combine $-2/3$ obtained from $2/3p^2(p^2+m^2_2)$ with
$-1/3$ from $1/3p^2(p^2+m^2_0)$.  Residues to each pole in
(\ref{mamp}) are $\{[2,-1],[-2,\frac{2}{3}],\frac{1}{3}\}$ from top
to bottom.

On the other hand,  the corresponding Newtonian potential energy
between two static sources
$\tilde{T}_{\mn}=M_1\delta^0_\mu\delta^0_\nu
\delta^3(\vec{x}-\vec{x}_1)$ and
$T^{\mn}=M_2\delta_0^\mu\delta_0^\nu \delta^3(\vec{x}-\vec{x}_2)$ is
given by \beq
U(r)=\frac{GM_1M_2}{r}\Bigg[-\frac{1}{r}+\frac{4}{3}\frac{e^{-m_2r}}{r}-\frac{1}{3}\frac{e^{-m_0r}}{r}\Bigg],~~r=|\vec{x}_1-\vec{x}_2|,
\eeq where the second term from ghost represents a repulsive force.
The factors are clearly understood when considering
$\tilde{T}_{\mn}T^{\mn}=\tilde{T}_{00}T_{00}$ and
$\tilde{T}T=\tilde{T}_{00}T_{00}$. This shows  why we started with
external source $T_{\mn}$.

Now we would like to mention the vDVZ discontinuity by pointing out
a factor of $\frac{2}{3}$ at the last term in the second line of
(\ref{mamp}). In this case, the external source is necessary to
compute residue and DOF.  In order to avoid the ghost issue, one
includes the Fierz-Pauli term instead of higher curvature terms in
the bilinear Lagrangian as \beq \label{fpterm}I_{\rm
FP}=-\frac{M^2_{\rm FP}}{4\kappa^2}\int d^4x
\sqrt{-\bar{g}}\Big(h_{\mn}^2-h^2\Big), \eeq  its amplitude leads to
when choosing the gauge condition  \beq \label{fpamp}[4A(p)]_{\rm
FP}= 2\Bigg[\tilde{T}_{\mn}\frac{1}{p^2+M_{\rm FP}^2}
T_{\mn}-\frac{1}{3} \tilde{T}\frac{1}{p^2+M_{\rm FP}^2} T\Bigg] \eeq
neglecting all contact terms (see (2.92) in Ref \cite{blas}). We
call the former (latter) as the tensor (scalar) amplitudes,
respectively.  Residues to each pole in (\ref{fpamp}) are
$[2,-\frac{2}{3}]$, which shows that the Fierz-Pauli massive gravity
is free from the ghost. The residue ``$-\frac{2}{3}$" in
(\ref{fpamp}) differs from ``$-1$" in
 the first line of (\ref{mamp}), producing
the famous vDVZ discontinuity in the massless limit of $M^2_{\rm FP}
\to 0$.  In general, this reflects a decreasing DOF of $5 \to 3$ in
the limit of $M^2_{\rm FP} \to 0$.  The same happens when replacing
$M^2_{\rm FP}$ by $m^2_2$ in the in the second line of (\ref{mamp})
even it shows  ghost states.

In order to investigate what happens in the massless limit of
$M^2_{\rm FP} \to0$ explicitly, we introduce the ADM formalism where
the metric is parameterized as \beq ds_{\rm ADM}^2= - N^2  dt^2 +
g_{ij} \Big(dx^i - N^i dt\Big) \Big(dx^j - N^j dt\Big). \eeq  Then,
we have to consider the noncovariant perturbation around the
Minkowski space as~\cite{Rub,Myungm} \beq \label{decom1} g_{ij}=
\delta_{ij}+ h_{ij},~N= 1+  n,~N_i=  n_i \eeq with \beqa
\label{pert}
 n =-\frac{1}{2}A,~
 n_i=\partial_iB+V_i,~
 h_{ij}=\psi\delta_{ij}+\partial_i\partial_j E+2\partial_{(i}F_{j)}+t_{ij}.\label{decom2} \eeqa
Here the conditions of
$\partial^iF_i=\partial^iV_i=\partial^it_{ij}=t_{ii}=0$ are imposed.
 The
last two conditions mean that $t_{ij}$ is a transverse and traceless
tensor in three spatial dimensions.  Using this decomposition, four
scalar modes ($A,B,\psi,E$), two vector modes ($V_i,F_i$), and a
tensor mode ($t_{ij}$) decouple completely from each other. These
all amount to 10 DOF for a symmetric tensor in four dimensions.
Considering the Fierz-Pauli mass term (\ref{fpterm}),  we show that
out of the 5 DOF of a massive graviton, 2 of these are expressed as
transverse and traceless tensor modes $t_{ij}$, 2 of these are
expressed as transverse vector modes $w_i=V_i-\dot{F}_i$, and the
remaining one is from a scalar $\psi$.  We  introduce the external
source term \beq \label{extsou} S_{int}=-\frac{1}{2\kappa^2}\int dt
d^3x\Big[h^{ij}T_{ij}+2h^{0j}T_{0j}+h^{00}T_{00}\Big] \eeq which
leads to \beq \label{gisource} S_{int}=-\frac{1}{2\kappa^2}\int dt
d^3x\Big[t_{ij}T_{ij}-2w_iT_{0i}+ \Phi T_{00}+\psi T_{ii}\Big] \eeq
expressed in terms of gauge-invariant modes of $\psi,
~w_i=V_i-\dot{F}_i,$ and $\Phi=A-2\dot{B}+\ddot{E}$. Here $\Phi$
plays the role of Newtonian potential.  The tensor equation takes a
relatively simple form \beq \label{tensors}
 t_{ij}=\frac{1}{\nabla^2-M^2_{\rm FP}}T_{ij}. \eeq We find that
there is no the vDVZ discontinuity in the massless limit because a
single mass term $M^2_{\rm FP}$ is present. This means that $t_{ij}$
describes 2DOF in the massless limit as \beq \label{mltensor}
 t_{ij}=\frac{1}{\nabla^2}T_{ij}. \eeq
  In the massless limit, a gauge-invariant vector reduces to \beq
\label{vecfe}
w_i=\frac{1}{\partial^2}T_{0i},~\partial^2=\partial_i\partial^i \eeq
which shows that the vector is a non-propagating mode because
temporal derivative term $\partial_0^2$ is absent here. Finally, in
the massless limit, $\psi$ leads to \beq
\psi=\frac{T_{ii}+2T_{00}-\frac{3\ddot{T}_{00}}{\partial^2}}{6\nabla^2},\eeq
which takes a further  form  \beq \label{fpt}
\psi=\frac{T_{00}}{2\partial^2}+\frac{T_{ii}-T_{00}}{6\nabla^2}.
\eeq This confirms the presence of the vDVZ discontinuity of the
Fierz-Pauli mass  term because the last term in (\ref{fpt})  implies
that $\psi$ is a propagating degree of freedom.

Consequently, in the massless limit of $M^2_{\rm FP} \to 0$, total
DOF is 3(=2+1) from $t_{ij}$ and $\psi$,  whereas  $\psi$ is absent
in the GR. This shows the vDVZ discontinuity unambiguously in the
Minkowski space. This is important because the extra scalar $\psi$
predicts a value for the gravitational bending of light by a massive
source that is $3/4$ of the Einstein prediction~\cite{POR}.

\section{Residue calculation for $\alpha=-3\beta$}

Choosing $\alpha=-3\beta$ for the non-critical gravity,  the
expression of (\ref{opamp}) reduces to \beqa
\label{cramp}[4A]_{\alpha=-3\beta} =&-&
2m^2_2\tilde{T}_{\mn}\frac{1}{\Big(\bar{\nabla}^2-\frac{4\Lambda}{3}-m^2_2
\Big)(\Delta_L^{(2)}-2\Lambda)}T^{\mn}
\nonumber \\
 &-&\frac{2m^2_2}{3}\tilde{T}\frac{1}{\Big(\bar{\nabla}^2-\frac{4\Lambda}{3}-m^2_2
\Big)(\bar{\nabla}^2+2\Lambda)}T
\\ \nonumber
&+&\frac{2m^2_2\Lambda}{9}\tilde{T}\frac{1}{\Big(\bar{\nabla}^2-\frac{4\Lambda}{3}-m^2_2
\Big)(\bar{\nabla}^2+2\Lambda)(\bar{\nabla}^2+\frac{4\Lambda}{3})}T
\\ \nonumber
&+&
\frac{1}{3}\tilde{T}\frac{1}{\bar{\nabla}^2+\frac{4\Lambda}{3}}T.
\eeqa
 For $\Lambda\not=0$, this is quite a non-trivial integral.
Hence, we can study the particle spectrum of graviton and scalar by
investigating  each  pole structure of their amplitude.  Here we
read off three scalar poles ($\tilde{T}T$) from (\ref{cramp}). Given
these poles, finding their residues is easy.  We wish to compute the
residue at each pole.

(a) Pole at $\bar{\nabla}^2=-\frac{4\Lambda}{3}$

\noindent The last two terms of (\ref{cramp}) contribute to the
scalar  residue at this pole as \beq
-\Bigg[\frac{8\Lambda}{3m^2_2+8\Lambda}\Bigg]\tilde{T}T. \eeq The
case of $\Lambda=0$  leads to the zero residue. As was mentioned
before, this corresponds to an unphysical pole.

(b) Pole at $\bar{\nabla}^2=-2\Lambda$

\noindent The scalar residue at this physical pole takes the form
\beq \label{res11}  -\Bigg[\frac{3m^2_2}{3m^2_2+10
 \Lambda}\Bigg]\tilde{T}T. \eeq
As was pointed out previously, this pole may contain the
gravitational waves in the Minkowski space limit for the Fierz-Pauli
massive gravity.  In order to recover it, we may take  the limit of
$\Lambda \to 0$. In this limit,  the scalar residue is $-1$, and
thus, the amplitude describes a massless graviton with 2 DOF like
\beq \label{massl-amp}\lim_{\Lambda\to 0}
\Big[4A\Big]_{\alpha=-3\beta}=2\Bigg[
\tilde{T}_{\mn}\frac{1}{p^2}T^{\mn}-\frac{1}{2}
\tilde{T}\frac{1}{p^2}T\Bigg], \eeq which is the first line of
(\ref{mamp}).

 (c) Pole at
$\bar{\nabla}^2=\frac{4\Lambda}{3}+m^2_2$

\noindent This is a new massive pole due to $\alpha
R^{\mn}R_{\mn}$-term. Its scalar residue takes the form \beq
\Bigg[\frac{2m^2_2(3m^2_2+7 \Lambda)}{(3m^2_2+10\Lambda)(3m^2_2+8
\Lambda)}\Bigg]. \eeq In the limit of $\Lambda \to 0$, it leads to
$2/3$.  However, one could not calculate the tensor residue unless
one knows $\Delta^{(2)}_L$ explicitly. In order to compute it, one
requires a further condition in the next section. This is a
traceless condition of $h=0$.

\section{Residue with $\alpha=-3\beta$ and TT gauge}
Requiring the traceless condition, we are working with the TT gauge.
In this case, one finds that \beq
\Delta^{(2)}_Lh_{\mn}=-\bar{\nabla}^2h_{\mn}+\frac{8\Lambda}{3}h_{\mn}.
\eeq The scattering amplitude takes the form \beqa
\label{ttamp}[4A]^{TT}_{\alpha=-3\beta}
=\tilde{T}_{\mn}h^{TT}_{\mn}=&&
2m^2_2\tilde{T}_{\mn}\frac{1}{\Big(\bar{\nabla}^2-\frac{4\Lambda}{3}-m^2_2
\Big)\Big(\bar{\nabla}^2-\frac{2\Lambda}{3}\Big)}T^{\mn}
\nonumber \\
 &-&\frac{2m^2_2}{3}\tilde{T}\frac{1}{\Big(\bar{\nabla}^2-\frac{4\Lambda}{3}-m^2_2
\Big)(\bar{\nabla}^2+2\Lambda)}T
\\ \nonumber
&+&\frac{2m^2_2\Lambda}{9}\tilde{T}\frac{1}{\Big(\bar{\nabla}^2-\frac{4\Lambda}{3}-m^2_2
\Big)(\bar{\nabla}^2+2\Lambda)(\bar{\nabla}^2+\frac{4\Lambda}{3})}T.
\eeqa Here are two tensor poles ($\tilde{T}_{\mn}T^{\mn}$) and four
different scalar poles ($\tilde{T}T$).

 (a) Pole at
$\bar{\nabla}^2=\frac{4\Lambda}{3}+m^2_2$

\noindent Comparing this pole with the AdS$_4$ pole
($\bar{\nabla}^2-\frac{2\Lambda}{3}$), it requires an inequality for
$m^2_2$ as~\cite{LP} \beq 0<m_2^2 < -\frac{2\Lambda}{3}, \eeq where
the saturation of right bound implies the critical gravity. This
bound for non-critical gravity is necessary to discuss the vDVZ
discontinuity in AdS$_4$ space. Similarly, one has a bound of
$0<M^2_{\rm FP} <2\Lambda/3$ in the dS$_4$ space~\cite{Hig}.  We
note that the critical gravity of $m^2_2=-2\Lambda/3$ is beyond the
above bound. Its residue is calculated to be \beq
-\Bigg[\frac{2m^2_2}{m^2_2+\frac{2
\Lambda}{3}}\Bigg]\tilde{T}_{\mn}T^{\mn}+\Bigg[\frac{2m^2_2(3m^2_2+7
\Lambda)}{(3m^2_2+10\Lambda)(3m^2_2+8 \Lambda)}\Bigg]\tilde{T}T.
\eeq  Taking first the $m^2_2\to 0$ limit and then, the $\Lambda \to
0$ limit leads to $[0,0]$, which means the zero residues.  However,
one has the scalar residue \beq
\label{refp}\Bigg[\frac{2\Lambda-2M^2_{\rm FP}}{3M^2_{\rm
FP}-2\Lambda}\Bigg]\tilde{T}T \eeq for the Fierz-Pauli massive
gravity in AdS$_4$ space~\cite{POR}. In the limit of the $M^2_{\rm
FP} \to0$ first, it leads to $-1$ (no vDVZ discontinuity), while for
the $\Lambda \to 0$ limit first, it takes a form of residue $-2/3$
(vDVZ discontinuity). Hence, this case of zero residue  is special
when comparing with the Fierz-Pauli massive gravity in AdS$_4$
space.

 On the other hand, taking first the
$\Lambda \to 0$ limit and then, the  $m^2_2\to 0$ limit leads to the
residues of $[-2,\frac{2}{3}]$, which means that the vDVZ
discontinuity occurs  in the ghost expression. Especially,  at the
critical point of $m^2_2=-2\Lambda/3$, the first term blows up
($-\infty$), while the second term gives us its residue of $-5/36$.
This means that it is impossible for the critical gravity to have a
finite tensor residue.  This is a new feature of residue of
$\frac{1}{p^2-\frac{4\Lambda}{3}+m^2_2}$  pole at the critical
point.

(b) Pole at $\bar{\nabla}^2=\frac{2\Lambda}{3}$

\noindent The tensor  residue is calculated to be

\beq \Bigg[\frac{2m^2_2}{m^2_2+\frac{2
\Lambda}{3}}\Bigg]\tilde{T}_{\mn}T^{\mn}, \eeq which blows up
($\infty$) for the critical case of $m^2_2=-2\Lambda/3$. This can be
confirmed by rewriting the first line of (\ref{ttamp}) as \beq
[4A]^{TT}_{(\alpha=-3\beta)} \to \Bigg[\frac{2m^2_2}{m^2_2+\frac{2
\Lambda}{3}}\Bigg]\tilde{T}_{\mn}\Bigg[\frac{1}{\bar{\nabla}^2-\frac{4\Lambda}{3}-m^2_2}
-\frac{1}{\bar{\nabla}^2-\frac{2\Lambda}{3}}\Bigg]T^{\mn}.\eeq For
the residue calculation, its momentum space representation takes the
form \beq \label{psamp} [4A(p)]^{TT}_{(\alpha=-3\beta)} \to
\Bigg[\frac{2m^2_2}{m^2_2+\frac{2
\Lambda}{3}}\Bigg]\tilde{T}_{\mn}\Bigg[-\frac{1}{p^2+\frac{4\Lambda}{3}+m^2_2}
+\frac{1}{p^2+\frac{2\Lambda}{3}}\Bigg]T^{\mn}.\eeq This shows a
massive graviton with negative energy (ghost).

c) Pole at $\bar{\nabla}^2=-\frac{4\Lambda}{3}$

\noindent The scalar residue at this  pole is \beq
\Bigg[\frac{m^2_2}{3m^2_2+8\Lambda}\Bigg] \tilde{T}T\eeq which is
contributed from the last term of (\ref{ttamp}) only. The case of
$\Lambda=0$ shows its residue of  $1/3$. This corresponds to an
unphysical pole.

(d) Pole at $\bar{\nabla}^2=-2\Lambda$

\noindent The scalar residue at this  pole takes the form \beq
\label{res1} - \Bigg[ \frac{3m^2_2}{3m^2_2+10
 \Lambda}\Bigg]\tilde{T}T. \eeq
This pole may contain the gravitational waves in the Minkowski space
limit. In order to recover it, we may take  the limit of $\Lambda
\to 0$. In this limit,  the residue is $-1$, and thus, the amplitude
describes a massless graviton with 2 DOF like as in  the first line
of (\ref{mamp}).

On the other hand, without external source,  equation (\ref{lineq})
satisfies the fourth-order equation under the TT gauge~\cite{LP}
\beq \label{fourth}
\Big(\bar{\nabla}^2-\frac{4\Lambda}{3}-m^2_2\Big)\Big(\bar{\nabla}^2-\frac{2\Lambda}{3}\Big)h_{\mn}=0
\eeq which is obtained by taking into account  the first line of
(\ref{ttamp}) only.  Hence, one did  not consider the scalar
amplitudes of the last two terms. For $m^2_2\not=-2\Lambda/3$, it
implies  a massless graviton satisfying the second-order equation
\beq \label{massless}
\Big(\bar{\nabla}^2-\frac{2\Lambda}{3}\Big)h^{m}_{\mn}=0 \eeq and a
massive graviton satisfying  the second-order equation \beq
\label{massive}
\Big(\bar{\nabla}^2-\frac{4\Lambda}{3}-m^2_2\Big)h^{M}_{\mn}=0. \eeq
For $m^2_2=-2\Lambda/3$, (\ref{massive}) degenerates
(\ref{massless}), which defines the critical gravity. Precisely
speaking, the fourth-order equation (\ref{fourth}) cannot be split
into two second-order equations (\ref{massless}) and (\ref{massive})
for $m^2_2=-2\Lambda/3$. Its promising  form is the fourth-order
equation \beq
\label{critical4}\Big(\bar{\nabla}^2-\frac{2\Lambda}{3}\Big)^2h^{\rm
cr}_{\mn}=0 \eeq for the critical gravity. The solution is given by
the log-gravity~\cite{BHRT} \beq h^{\rm cr}_{\mn}=h^{\rm
log}_{\mn}=f(t,\rho)h^{m}_{\mn} \eeq where \beq f(t,\rho)=2it
+\ln[\sinh(2\rho)]-\ln[\tanh \rho].  \eeq However, the non-unitarity
issue of log-gravity is not still resolved, indicating that the log
gravity suffers from the ghost problem~\cite{PR,LLL}.

\section{Discussions}

We start by noting that in the  FP massive gravity, its scalar
residue of $\frac{1}{p^2+M_{\rm FP}^2}$ pole is $-2/3$ in the
Minkowski space, showing the vDVZ discontinuity instead of $-1$ for
the massless case. In the AdS$_4$ space, its scalar residue of
$\frac{1}{p^2-2\Lambda +M^2_{\rm FP}} $ pole in (\ref{refp}) is $-1$
for taking the $M^2_{\rm FP} \to 0$ limit first, while its scalar
residue is given by $-2/3$ for taking the $\Lambda \to 0$ limit
first.

In the higher curvature gravity, the scalar residue of
$\frac{1}{p^2+m^2_2}$ pole  is $2/3$ in the Minkowski space because
of the ghost state. On the other hand, for non-critical gravity with
$0<m^2_2<-2\Lambda/3$,  the scalar residue of
$\frac{1}{p^2+\frac{4\Lambda}{3}+m^2_2}$ pole is $0$ for taking the
$m^2_2 \to 0$ limit first, while its scalar residue is $2/3$ for
taking the $\Lambda \to 0$ limit first because of the ghost state.
The latter represents that propagating DOF is decreased from 5 to 3
in the massless limit of $m^2_2 \to 0$.

Considering that the vDVZ discontinuity is the massless limit of the
massive gravity in the Minkowski space,   the critical gravity
appears as the massless limit of higher curvature gravity in the
AdS$_4$ space.  Hence, we proposed that the critical gravity may be
represented as the vDVZ discontinuity of the  higher curvature
gravity in the AdS$_4$ space.  In this case, we expect to have  a
decreasing DOF ($5 \to 3$)  at the critical point. However, at the
critical point of $m^2_2=-2\Lambda/3$, the tensor residue of
$\frac{1}{p^2+\frac{4\Lambda}{3}+m^2_2}$ blows up, while its scalar
residue is $-5/36$. This is surely an unpromising feature of the
critical gravity when plugging external sources.

Finally, we wish to comment on how to understand the log-gravity
solution at the critical point when plugging the external source. In
our approach, the conditions of transverse and traceless were taken
into account by the external source, while these conditions were
used to derive (\ref{critical4}) and,  obtain the massless modes
$h^{m}_{\mn}$ and its log-modes $h^{\rm log}_{\mn}$.  Even though
the log-modes has positive excitation energy, it was pointed out
that negative norm states (ghost) appears from linear combination of
massless and log-modes~\cite{PR}. Eventually, the ghost-free
condition requires the truncation of log-modes by choosing an
appropriate boundary condition~\cite{LP,LPP}. However, this means
that all massive modes are eliminated from the IR region, leaving
with the Einstein gravity.  In this case, the role of higher
curvature terms is unclear at the linearized level.

\section*{Acknowledgment}
 This work was supported by the National Research
Foundation of Korea (NRF) grant funded by the Korea government
(MEST) (No.2010-0028080).

\end{document}